\documentclass{ws-mpla}

\begin{document}

\markboth{Petre Di\c t\u a}
{  Unitarity 
condition method versus standard unitarity triangle approach}

%%%%%%%%%%%%%%%%%%%%% Publisher's Area please ignore %%%%%%%%%%%%%%%
%
\catchline{}{}{}{}{}
%
%%%%%%%%%%%%%%%%%%%%%%%%%%%%%%%%%%%%%%%%%%%%%%%%%%%%%%%%%%%%%%%%%%%%

\title{\bf GLOBAL FITS TO THE CABIBBO-KOBAYASHI-MASKAWA MATRIX: UNITARITY CONDITION METHOD VERSUS STANDARD UNITARITY TRIANGLES APPROACH}

\author{\footnotesize PETRE DITA}

\address{Institute of Physics and Nuclear Engineering,\\
P.O. Box MG6, Bucharest, Romania  \\
dita@zeus.theory.nipne.ro }

\maketitle

\pub{Received (Day Month Year)}{Revised (Day Month Year)}

\begin{abstract}

The aim of the paper is to make a comparison between the unitarity condition method and the standard version of the unitarity triangle approach by using as parameters four independent moduli $|U_{ij}|$. This choice is motivated by the measurability property and leads to a simple criterion for the separation of unistochastic matrices from the double stochastic ones, whose fulfillment is the key point of any global fit for the CKM entries. In our formulation both the methods are exact, do not depend on any assumption as the smallness of some parameter and both can be used to global fits in the quark and fermion sectors. Monte Carlo simulations show that the separation criterion puts very strong conditions requiring a  fine tuning of all the CKM matrix elements.
\keywords{CP violation, global fits,unitarity condition method }
\end{abstract}

\ccode{PACS Nos.: 12.15 -y, 12.15.Hh}

\section{Introduction }	
 The Cabibbo-Kobayashi-Maskawa (CKM)  matrix  is a lively subject in particle physics and  the determination of its entries  that govern all flavor changing transitions of quarks and leptons in the Standard Model is an
important task for both experimenters and theorists. The large interest in the subject is also reflected by the workshops organized in the last  years whose main subject was the CKM matrix$^{1-4}$, one of the main goals of these gatherings being global fits to the CKM matrix parameters. The usual method for doing that is the so called standard unitarity triangle approach  that make use of the orthogonality property between the first column and the third one of
 the  recommended form for the CKM matrix, see Refs. 5-8. 

 Recently\cite{Di}, we have proposed  a unitarity condition method for  constraining the CKM matrix entries, method that fully exploits the unitarity properties of CKM matrix, it is  an  exact one,  its use does not depend on any approximation, and can be applied to both  quark and fermion sectors.

The main goal of the paper is to make a comparison between the two approaches alluded to the title, and for doing that we  reformulate the standard unitarity triangle method by requiring that the triangles sides have to depend on 
four  measurable quantities. The measurability property was first raised by  
 Jarlskog$^{10-12}$ who has shown that in the $3$-flavors case only the functions that depend on the  moduli $|U_{ij}|$ and/or the ``angular looking objects'' $U_{\alpha j}U_{\beta k}U^*_{\alpha k}U^*_{\beta j}$ are re-phasing invariant and {\em measurable}.
%\end{document}
 We remark that the common {\em four} independent parameters used in the standard unitarity triangle approach are $|U_{us}|,\,|U_{ub}|,\,\,|U_{cb}|,$ and the phase $\delta$, see Refs. 6-8 and the references cited therein, where the notations are the well-known ones. Concerning  $\delta$ it is not seen as being  directly measurable, at least in a single-type experiment, even it enters in  a measurable quantity, the Jarlskog invariant 
$J=Im(U_{\alpha j}U_{\beta k}U^*_{\alpha k}U^*_{\beta j})$\cite{Ja1}. The $J$ expression suggests that the ``natural'' four independent parameters  
to be used in any global fit  have to be the moduli. Thus in the following we will always use as parameters {\em four} independent moduli $|U_{ij}|$, and in particular we will provide another measurable expression  for $\delta$. That choice allows us to find a simple criterion for the separation of double stochastic matrices from the unistochastic ones, problem that to my knowledge  was not yet considered  in the physical literature, and its fulfillment is the key point in any global fit of the CKM entries.

The paper is organized as follows. In the next section we find the gauge group of unitary matrices, i.e. the group of elementary transformations  whose action on the  unitary matrices does not change their unitary properties and physical content. In Sec. 3, we expose the unitarity condition method\cite{Di,Di1} and provide the necessary and sufficient conditions for discrimination between the doubly stochastic matrices and the unitary matrices. We show that the unitarity condition is a very strong one and doing a global fit is not an easy matter. In Sec. 4, we reformulate the standard unitarity triangle approach$^{5-8}$ by using in all formulas only measurable quantities. In Sec. 5, we present a comparison between the two approaches,  and we conclude in Sec. 6.

\section {Unitary matrices and their gauge group }

The  $n$-generations CKM matrix is assumed to  be unitary,  and
it  depends on $n^2$ parameters which are usually  taken as $n(n-1)/2$ angles and $n(n+1)/2$ phases, each set  taking values in $[0,\pi/2]$, and respectively, $[0,2 \pi)$. Very soon\cite{Ma} it was realized that by a redefinition of  quark fields, the elements of the first row $3\times 3$ unitary matrix can be made positive. That observation was formalized as follows: the  ``encoded'' physics  in the CKM unitary matrix is invariant to multiplication at right and/or left by diagonal phase matrices $D=$diagonal$(e^{i\psi_1},\dots,e^{i\psi_n})$, see e.g. Refs. 10, 15 and 16, because we can redefine the quark fields phases. That means that  we have at our disposal 
$2n-1$ phases whose values can be chosen by us, and the common choice is   $0$ and/or $\pi$. For a non-standard choice in the $3$-dimensional case see Ref 17. That leads to $n(n-1)/2 -(2n-1)=(n-1)(n-2)/2$ independent phases and 
$n(n-1)/2$ independent angles that parametrize a unitary matrix. Here we rise  a novel problem: are there other  {\em elementary} transformations  of unitary matrices under which their  physical  properties  do not change?. 

Besides the left and/or right multiplication by diagonal phase matrices there is another transformation: multiplication at left and/or right by permutation matrices. Permutation matrices are matrices whose elements on each row and each column are zero, but one that equals unity. They interchange columns and, respectively, rows between themselves. Both the diagonal phase matrices and permutation matrices are subgroups of  unitary matrices. If $D$ is a diagonal phase matrix and $P$ a permutation matrix then 
\[D D^*=P P^*=I_n\]
where $*$ denotes the adjoint, and $I_n$ is the $n$-dimensional unit matrix. 
Another equivalent unitary matrix can be obtained by taking the transpose of the original one, and if we would require that both the  matrices should have similar forms after such a transformation,  the Kobayashi-Maskawa form\cite{KM} will be the compulsory 
choice, and not the usual one\cite{CK}. We can also apply the complex conjugation to all its entries, and all its physical properties do not change. If we denote by $T$ the transpose and by $C$ the complex conjugation, these transformations form a subgroup of each other because
\[T^2=C^2=Identity\]
In conclusion the product group 
\begin{eqnarray}
G=D \times P \times T\times C\label{gauge}\end{eqnarray}
is the gauge invariance group of unitary matrices, i.e.  the physical content remains invariant by applying a number of the above transformations to a given unitary matrix. The $C$-invariance has an important consequence: the range of all the $(n-1)(n-2)/2$ independent phases is $[0,\pi]$.

\section{Unitarity condition method}

 In the following we  make use of the rephasing invariance property and multiply the recommended form for CKM matrix\cite{PDG2004} at left and, respectively, at right by the diagonal phase matrices $D_l=(1,e^{-i \delta},e^{-i \delta})$ and $D_r=(1,1,e^{i \delta})$, and obtain the form that will used in the following
\begin{eqnarray}
U=
\left(\begin{array}{ccc}
c_{12}c_{13}&c_{13}s_{12}&s_{13}\\
-c_{23}s_{12}e^{-i \delta}-c_{12}s_{23}s_{13}&c_{12}c_{23}e^{-i \delta}-s_{12}s_{23}s_{13}&s_{23}c_{13}\\
s_{12}s_{23}e^{-i\delta}-c_{12}c_{23}s_{13}&-c_{12}s_{23}e^{-i \delta}-s_{12}c_{23}s_{13}&c_{23}c_{13}
\end{array}\right) \label{ckm}
\end{eqnarray}
with $c_{ij}=\cos \theta_{ij}$ and  $s_{ij}=\sin \theta_{ij}$
for the generation labels $ij=12, 13, 23$, and $\delta$ is the phase that encodes the breaking of the $CP$-invariance. 
The above form depends on {\em four} independent phases, the phases of $U_{21},\,U_{22},\,  U_{31},\,U_{32} $, instead of {\em fives} in the usual form. These four phases can be considered as four fundamental frequencies, and their knowledge is equivalent to the knowledge of all the $U$ entries, see e.g. Ref. 21.

The above model has to be supplemented by experimental data 
that are usually supplied by experimenters under the form of a positive entries matrix 
\begin{eqnarray}
V=\left(\begin{array}{ccc}
\vspace*{1mm}
V_{ud}^2&V_{us}^2&V_{ub}^2\\
\vspace*{1mm}
V_{cd}^2&V_{cs}^2&V_{cb}^2\\

V_{td}^2&V_{ts}^2&V_{tb}^2\\
\end{array}\right)\label{pos}
\end{eqnarray}
More generally, the experimental data are given in terms of some functions $f_k(V_{ij}),\,k=1,\dots,N$, that depend on the $V$ entries. For defining a theoretical model we need also
a linking relation between the above two objects, and we propose
\begin{eqnarray}  V=|U|^2\label{rel}\end{eqnarray}
The previous equation is equivalent to the following relations
\begin{eqnarray}%\begin{array}{l}
V_{ud}^2&=&c^2_{12} c^2_{13},\,\, V_{us}^2=s^2_{12}c^2_{13},\,\,V_{ub}^2=s^2_{13}\nonumber \\
 V_{cb}^2&=&s^2_{23} c^2_{13},\,\,
 V_{tb}^2=c^2_{13} c^2_{23},\nonumber\\
V_{cd}^2&=&s^2_{12} c^2_{23}+s^2_{13} s^2_{23} c^2_{12}+2 s_{12}s_{13}s_{23}c_{12}c_{23}\cos\delta,\nonumber\\
V_{cs}^2&=&c^2_{12} c^2_{23}+s^2_{12} s^2_{13} s^2_{23}-2 s_{12}s_{13}s_{23}c_{12}c_{23}\cos\delta\nonumber,\\
V_{td}^2&=&s^2_{13}c^2_{12}c^2_{23}+s^2_{12}s^2_{23}-2 s_{12}s_{13}s_{23}c_{12}c_{23}\cos\delta\nonumber,\\
V_{ts}^2&=&s^2_{12} s^2_{13} c^2_{23}+c^2_{12}s^2_{23} +2 s_{12}s_{13}s_{23}c_{12}c_{23}\cos\delta \label{unitary}
\end{eqnarray}
The main theoretical problem is to see if from a matrix as (\ref{pos}) one can reconstruct a unitary matrix as (\ref{ckm}). For that
we make use of the unitarity relation
\begin{eqnarray}  U U^* = U^* U = I_3\label{unitarity}\end{eqnarray}
to obtain from the relations (\ref{unitary}) the {\em weakest} form of  unitarity
\begin{eqnarray}
\sum_{i=d,s,b} V_{ji}^2-1=0, \quad j=u,c,t\nonumber \\
\sum_{i=u,c,t} V_{ij}^2-1=0, \quad j=d,s,b\label{sto}
\end{eqnarray}
For the moment we assume that the entries of $V$ are such that the  relations (\ref{sto}) are exactly satisfied; then the set (\ref{sto}) is known as {\em\ double stochastic matrices}, and the  subset  $V_{ij}^2=|U_{ij}|^2$ is known as {\em unistochastic matrices}\cite{MO}. The constraints  (\ref{sto}) are necessary but not sufficient for unitarity, where from we infer that we need a separation criterion between the two sets. 

Furthermore, for simplicity we will  choose four independent parameters to work with, and our choice is 
\begin{eqnarray} |U_{us}|=a,\,|U_{ub}|=b,\,|U_{cd}|=d,\,|U_{cb}|=c \label{choice}\end{eqnarray}
Using the exact doubly stochasticity of $V$ we form the real matrix
\begin{eqnarray}S= \left(\begin{array}{ccc}
\vspace*{2mm}
\sqrt{1-a^2-b^2}&a&b\\
\vspace*{2mm}
 d&\sqrt{1-c^2-d^2}&c\\
\vspace*{2mm}
 \sqrt{a^2+b^2-d^2}&\sqrt{c^2+d^2-a^2}&\sqrt{1-b^2-c^2}
\end{array}\right)\label{dbl}
\end{eqnarray}
and observe that $|S|^2$ is double stochastic.

Making the identification $S_{ij}=V_{ij}$ in relation (\ref{unitary}) we find  expressions for $s_{ij}$ and  $\cos\delta$\cite{Di1}, the last one being given by
\begin{eqnarray}\cos\delta=\frac{d^2(1-b^2)^2-a^2(1-b^2-c^2)-b^2c^2(1-a^2-b^2)}{2 a b c \sqrt{1-a^2-b^2} \sqrt{1-b^2-c^2}}\label{cosd}\end{eqnarray}
showing that $\cos\delta$ is a measurable quantity being  expressed in terms of four moduli. 
The separation criterion between double stochastic and unistochastic matrices is given by the physical condition
\begin{eqnarray} -1\le \cos\delta\le 1,\qquad {\rm or}\,\,\, \cos^2\delta \le 1\label{unit}\end{eqnarray}
that is equivalent to
\begin{eqnarray}\begin{array}{c}
a^4-2 a^4 c^2-2 a^2 b^2 c^2+a^4 c^4+2 a^2 b^2 c^4+ b^4 c^4-2 a^2 d^2+2 a^2 b^2 d^2\\+2 a^2 c^2 d^2 -2 b^2 c^2 d^2
+2 a^2 b^2 c^2 d^2+2  b^4 c^2 d^2+d^4-2 b^2 d^4+ b^4 d^4 \leq 0 \end{array}
\end{eqnarray}
relation that describes the physically admissible region in the 4-dimensional space generated by the moduli $a,b,c,d$. This result shows that four independent moduli do not always determine a unitary matrix, even when the moduli satisfy the relations (\ref{sto}); this happens then and only then, when the relation (\ref{unit}) is satisfied. The above relation separates the unitary matrices from the doubly stochastic ones in our choice of independent moduli. The form (\ref{cosd}) changes when we change the choice (\ref{choice}).
Relations that have to be satisfied for fulfilling  unitarity have  also  been  found in Ref. 23 but unfortunately they were not taken into account by the CKM fitter community.

Relations (\ref{sto}) being exactly satisfied we can  compute $\cos\delta$
and check if the physical condition (\ref{unit}) is satisfied. If it is, then the experimental data are compatible with the form (\ref{ckm}) and we can easily reconstruct it from the data (\ref{pos}). If  the physical condition is violated there is no compatibility and the story ends here.

However, the experimental data are not known with infinite precision to see if 
the relations (\ref{pos}) are exactly satisfied. The data being affected by errors, what  could we do in that situation? There is the place where the gauge invariance group of unitary matrices enters the game.
The gauge invariance group (\ref{gauge}) tell us that there is a  ``democracy''  within the CKM matrix entries, all of them enjoying the same ``rights'', and by consequence  our choice freedom does not amount to the choice (\ref{choice}). 
It is a combinatorial problem to find all the {\em four} independent moduli groups, and in fact there are only 58 such  groups  that lead to 165 different expressions for  $\cos\delta$. Depending on the explicit group one gets one, two, three or four  different expressions for $\cos\delta$. The huge number for $\cos\delta$ is a consequence of the data errors and comes from the fact that neither relations (\ref{sto}) are exactly satisfied, nor $\cos\delta$ values are equal,  nor in the physical region; all these three conditions have to be simultaneously satisfied in order to find a unitary matrix. If the data are exactly known and satisfy the relations (\ref{sto}) all the 165 expressions give a single number. For example with the choice of rational numbers
\[a=\frac{11}{50}= 0.22,\,\, b=\frac{23}{6250}=0.00368,\,\, c=\frac{26}{625}=0.0416,\,\,
 d=\frac{28}{125}=0.224\]
 that are almost identical with the  values recommended in\cite{PDG2004}, the $S^2$  matrix  (\ref{dbl}) is doubly stochastic and one gets
\[\cos\delta=\frac{7748742441690187}{7187500\sqrt{ 144947356493366}}\approx 28.32\]
Although two conditions,  the double stochasticity property and the equality of $\cos\delta$ are exactly satisfied, the physical condition (\ref{unit}) is not, so the above numbers are far away from numbers coming from  a unitary matrix. If we change only one value and take $a=28/125=0.224$, we find $\cos\delta\approx 1.2897$. This simple test could say that the recommended value $a=0.22$ is highly improbable to be true and has to be changed. However we have to wait and see what the experiments will tell us!

 To see that the condition (\ref{unit}) is  constraining, we did a Monte Carlo simulation for
  (\ref{cosd}) using as input the PDG data $a=0.2200\pm0.0026,\,\,b=(3.67\pm 0.47)\times10^{-3},\,\ c=(41.3\pm 1.5)\times10^{-3},\,\, 
d=0.224\pm 0.012$, i.e. the values recommended in Ref. 20, and the results are shown in Fig.1. The results show that only a tiny fraction of simulated data are in the physical region and, by consequence, a global fit is not easy to be  done. For similar numerical results see Ref. 24.

\begin{figure}[th]
\centerline{\psfig{file=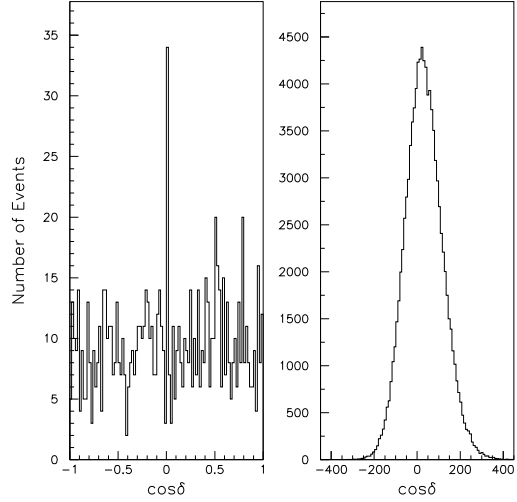,width=3.0in}}
\vspace*{8pt}
\caption{Number of Events as function of  $\cos\delta$. The input values are:
 $a=0.2200\pm0.0026,\,\,b=(3.67\pm 0.47)\times10^{-3},\,\ c=(41.3\pm 1.5)\times10^{-3},\,\, d=0.224\pm 0.012$. (a) On the left panel one sees the number of events for the physical  $\cos\delta\in (-1,1)$.   On the right panel are all the  events and the interval of variation is $\cos\delta\in[-350,\,\,450]$. The number of physical events is  $\approx 1\%$ of the total simulated data, and there is a peak around $\cos\delta\approx 0$.}
\end{figure}

In the case of approximate data the  explicit form for $\cos\delta$ depends on the independent four parameters  we choose to parameterize the data and also on the parameters entering the last four equations in (\ref{unitary}) that provide in general different expressions for $\cos\delta$, i.e.  the  numerical values are in general different  $\cos\delta^{(i)}\ne\cos\delta^{(j)},\,\, i\ne j $. In that situation we have to provide a $\chi^2-$test 
that has to take into account the double
stochasticity property expressed by the 
conditions (\ref{sto}) and the fact that in general the  data are such that
  $\cos\delta$ may take different numerical values even for those  $\cos\delta$  coming from 
a definite   choice of the four independent parameters, and more, the numerical values  could be  outside the physical range $(-1,1)$. My proposal is
\begin{eqnarray}
\chi^2_1=\sum_{i < j}(\cos\delta^{(i)} -\cos\delta^{(j)})^2+\sum_{j=u,c,t}\left(
\sum_{i=d,s,b}V_{ji}^2-1\right)^2\nonumber\\
+\sum_{j=d,s,b}\left(
\sum_{i=u,c,t}V_{ij}^2-1\right)^2,\,\,\,\,-1\le\cos\delta^{(i)}\le 1\label{chi} 
\end{eqnarray}
We stress that the  relation $\chi^2_1\equiv 0$ holds true for both double stochastic and unitary matrices, and only  the physical condition $-1\le\cos\delta^{(i)}\le 1$ discriminates between them such that it  has to be compulsory satisfied. The second component of the $\chi^2$-test 
has the form
\begin{eqnarray}
\chi^2_2=\sum_{i=1}\left(\frac{d_i-\widetilde{d_i}}{\sigma_i}\right)^2\label{chi1}\end{eqnarray}
where $d_i$ are theoretical functions depending on $V_{kl}$, $\widetilde{d_i}$ are the measured experimental data for $d_i$, and $\sigma$ is the vector of errors associated to $\widetilde{d}_{i}$. The formula  $\chi^2=\chi^2_1+\chi^2_2$ is our phenomenological tool for analyzing the experimental data.

\section{Standard unitarity triangle approach}

Unitarity triangle approach exploits another consequence of the unitarity property Eq.(\ref{unitarity}), namely the six  orthogonality relations of rows and,
respectively,  columns of a unitary matrix$^{6-8}$. Its root is in the pioneering work by Wolfenstein\cite{Wo} where he proposed a relationship between the theoretical object $U$ and the experimental data available then, its nice feature being the possibility to {\em estimate  the order of magnitude} for all $U$ entries at a time when the experimental data were very scarce.   Usually one considers only the orthogonality of the first and the third columns of $U$, relation which is written as
\begin{eqnarray}
U_{ud} U_{ub}^* + U_{cd} U_{cb}^* + U_{td} U_{tb}^*=0\label{ort}
\end{eqnarray}
The main reason for that choice is the structure  that arise in (\ref{ort}) by applying the Wolfenstein parametrization and keeping only the leading non-vanishing terms. However, there is another orthogonality relation that has the same structure, that given  by the orthogonality of first and third rows
\begin{eqnarray}
U_{ud} U_{td}^* + U_{us} U_{ts}^* + U_{ub} U_{tb}^*=0\label{ort1}
\end{eqnarray}
that was never used in  phenomenological analyzes. The above relations can be represented as triangles in the complex plane. The other four relations are neglected because the triangles have a flattened form in the complex plane.
To see how the relation (\ref{ort}) has been exploited till now, see Refs. 7, 8, 25 and 26, and the references cited therein.

The relations (\ref{ort})-(\ref{ort1}) are  scaled by dividing them  
  through the middle term such that the length of one side is 1. 
The other sides have the lengths

\begin{eqnarray}
R_u&=&\left|\frac{U_{ud} U_{ub}^*}{U_{cd} U_{cb}^*}\right|=\frac{b\,\sqrt{1-a^2-b^2}}{c\,d}\nonumber\\
R_t&=&\left|\frac{U_{td} U_{tb}^*}{U_{cd} U_{cb}^*}\right|=\frac{\sqrt{a^2+b^2-d^2}\,\,\sqrt{1-b^2-c^2}}{c\,d}\label{tri1}
\end{eqnarray}
for Eq.(\ref{ort}), and respectively 
\begin{eqnarray}
R^{'}_u&=&\left|\frac{U_{ud} U_{td}^*}{U_{us} U_{ts}^*}\right|=\frac{\sqrt{1-a^2-b^2}\,\sqrt{a^2+b^2-d^2}}{a\,\sqrt{c^2+d^2-a^2}}\nonumber\\
R^{'}_t&=&\left|\frac{U_{ub} U_{tb}^*}{U_{us} U_{ts}^*}\right|=\frac{b\,\sqrt{1-b^2-c^2}}{a\,\sqrt{c^2+d^2-a^2}}\label{tri3}
\end{eqnarray} for Eq.(\ref{ort1}).
 On the right hand side we have written the $R$-values in our choice for independent parameters (\ref{choice}) by supposing that the relations (\ref{sto}) hold. In this case we can compute the numerical values on the right hand side and  see if they form a triangle. If they do, one computes their angles and the above triangle provide us {\em two} independent re-phasing invariant angles. If the computed values  do not form a triangle, i.e. they do not satisfy the inequality
\begin{eqnarray}|R_u -R_ t|\le 1\le R_u+R_t\label{unit1}\end{eqnarray}
the data are not compatible to the existence of a unitary matrix. This inequality is the analogous condition to the physical condition (\ref{unit}). Monte Carlo simulations for $R_u$ and $R_t^2$ are shown in Fig.2. They tell us that there are no problems with $R_u \in [0.2,\,0.6]$. Concerning $R_t^2$, it takes values in the range $ [-250,\,250]$ and only $37.5\% $ are positive;  trying to satisfy the inequalities (\ref{unit1}) one finds that only $1.5 \%$ of them are in the physical region.
 The difficulties come from
$\sqrt{a^2+b^2-d^2}$ and $\sqrt{c^2+d^2-a^2}$,
 that lead easily to imaginary values  because $a\approx d$, and from the smallness of  $a,\,b,\,c$ and $d$. In the usual approach they are approximated by $|U_{td}|$, and, respectively, by $|U_{ts}|$, because $a^2+b^2-d^2=1-|U_{ud}|^2-d^2\approx |U_{td}|^2$, etc., such that the imaginary values are never seen.

\begin{figure}[th]
\centerline{\psfig{file=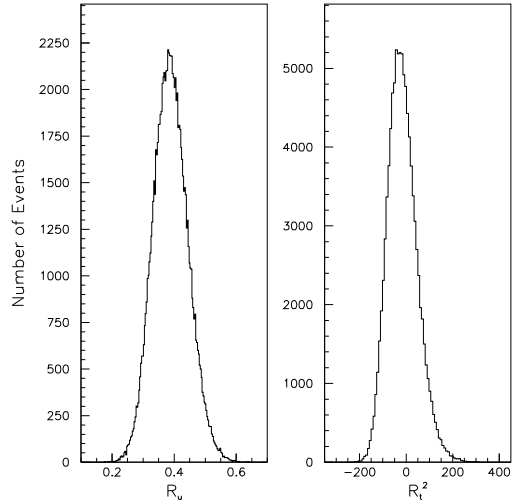,width=3.0in}}
\vspace*{8pt}
\caption{Number of Events as function of  $R_u$, and $R_t^2$. The input values are:
 $a=0.2200\pm0.0026,\,\,b=(3.67\pm 0.47)\times10^{-3},\,\ c=(41.3\pm 1.5)\times10^{-3},\,\, d=0.224\pm 0.012$. $R_u$ takes values within the interval  $R_u\in[0,2,\,\, 0.6]$, and $R_t^2\in [-250,\,\,250  ]$, and only $37.5\%$ of them are positive. The condition (\ref{unit1}) reduces the number of $R_t$ events to about $1.5\%$. }
\end{figure}

Similar to the unitarity condition method, when the data are not exactly known, we have to calculate all the $R_i$ forms corresponding to the 58 groups of four independent moduli. With the choice
\[|U_{ud}|=e,\,\, |U_{us}|=a,\,\,|U_{cd}|=d, \,\, |U_{cs}|=f\]
one gets

\begin{eqnarray}
R_u&=&\frac{e\,\sqrt{1-a^2-e^2}}{d\,\sqrt{1-d^2-f^2}}\nonumber\\
R_t&=&\frac{\sqrt{1-e^2-d^2}\sqrt{a^2+d^2+e^2+f^2-1}}{d\,\sqrt{1-d^2-f^2}}\label{tri2}
\end{eqnarray}
which is quite different from (\ref{tri1}), so we have to require $R_u^{(i)} \approx R_u^{(j)},\,\,i\ne j$ and respectively $R_t^{(i)} \approx R_t^{(j)},\,\,i\ne j$. Similar to the preceding case the problem  is caused by $\sqrt{1-d^2-f^2}$ that takes imaginary values because $f$ is quite large, and the determination of the two ratios (\ref{tri2}) is difficult.
 In this case of standard unitarity triangle  one can define a $\chi^2-$test as follows
\begin{eqnarray}
\chi^2_3&=&\sum_{i < j}(R_u^{(i)} -R_u^{(j)})^2+ (R_t^{(i)} -R_t^{(j)})^2
 +\sum_{j=u,c,t}\left(
\sum_{i=d,s,b}V_{ji}^2-1\right)^2 \nonumber \\&
+&\sum_{j=d,s,b}\left(
\sum_{i=u,c,t}V_{ij}^2-1\right)^2,\quad |R_u^{(i)} -R_ t^{(i)}|\le 1\le R_u^{(i)}+R_t^{(i)} \label{chi-3} 
\end{eqnarray}

Also we have to consider all the six orthogonality relations on the same footing and we have to include this information  in the form (\ref{chi-3}). The necessity for doing that comes from the fact that {\em one} orthogonality relation provides only {\em two} rephasing invariant angles. Or it is well known\cite{AKL} that there are necessary {\em four} such angles for a complete determination of a unitary matrix. Unfortunately the beautiful paper\cite{AKL} had no echo among the people working on global fits to the CKM matrix entries.

Similar to the preceding method
$\chi^2=\chi_3^2+\chi_2^2$ can be used as a $\chi^2$-test for the standard unitarity triangles approach. In principle both the approaches, unitarity condition method and  unitarity triangles method, have to give the same results modulo small errors. Since some fitting  methods could provide numerical values for all $V_{ij}$, the unitarity test $-1\le \cos\delta\le 1$ has to be checked.

\section{Comparison of the two methods}

Doing a global fit for the CKM matrix entries is not easy at all. The difficulty comes from the formulas (\ref{cosd}), (\ref{tri1}),(\ref{tri3}), (\ref{tri2}) and those similar to, that lead to unphysical values and we have to find numerically safe methods to satisfy the physical constraints (\ref{unit}) and/or (\ref{unit1}).Concerning $\delta$ the Monte Carlo simulations, see Fig.1, show that the overwhelming majority of its values lies outside the physical region (\ref{unit}).  For other choices of the independent parameters, $\cos\delta$ can even take  imaginary values  as happens when on computes it by using central $|U_{ij}|$ values  published in some fits\cite{Di}.  A similar situation arises for $R_u$ and/or $R_t$. Fig.2 shows that $R_t^2$ takes negative values in $62.5\%$ cases, i.e. $R_t$ gets imaginary. That happens because $a\approx d$. If we try to fulfill also the condition 
(\ref{unit1}) then the number of physical events is around $1.5\%$, i.e. of the same order of magnitude as for  $\cos\delta$. Perhaps it is a matter of taste to consider that it is numerically easier to satisfy the relation (\ref{unit}) and more difficult to implement the inequalities (\ref{unit1}); perhaps it comes from my   
dissatisfaction concerning the  current use of the unitarity triangles approach.

The main reason for looking for  an alternative approach to global CKM  fitting techniques is that 
 there is no simple relation between the parameters defined by Wolfenstein\cite{Wo}, $\lambda,\, A,\,\rho,\,\eta$, and directly measurable quantities. The weakness of Wolfenstein parametrization was intuited many years ago by Branco and Lavoura\cite{Br}, but, unfortunately, it was not fully exploited, although they gave a Wolfenstein-type parametrization with {\em four moduli!}. Their proposal was to choose another parameter, $q=|U_{td}/(U_{us}\,U_{cb})|^2$, instead of the usual one $\delta={\rm arg}(U_{ub}\,U_{cs}\,U_{us}^*\,U_{cb}^*)$, the first  being more relevant to $B-$mesons physics and a {\em measurable} quantity\cite{Br}. Of course, in principle, $\delta$ is measurable, see the relation (\ref{cosd}) that expresses  $\cos\delta$ through measurable quantities, but until now it cannot be measured in a {\em single} experiment; we have to use data from many different experiments to obtain its value. A similar situation is that of measuring the Jarlskog invariant $J$\cite{Ja1} that depends on $\sin\delta$.

As concerns the nowadays use of the unitarity triangle method\cite{Je,Bo}, in my opinion, its main drawback are  the  approximations  $\sqrt{a^2+b^2-d^2}\approx |U_{td}|$ and $\sqrt{c^2+d^2-a^2}\approx |U_{ts}|$, and similar to, for other choices of the independent parameters. In some happy situation as for example when the range of one $R_u$ is positive,  see e.g. the case (\ref{tri1}), one can ``control''  $R_t$  to obtain reasonable values for some physical quantities, even if $R_t$  can take imaginary values. However the problem is not simple and obtaining a physical result requires a very precise matching between the involved parameters. If we look at the triangle inequality (\ref{unit1}) we easily see that the equality $a=d$ is excluded since $R_u+R_t < 1$, for $a, b, {\rm and}\,\, c$ around the measured values. A physical solution is possible if their difference $a - d \sim 10^{-4} - 10^{-5}$, i.e. $a$ and $d$ are strongly correlated taking values in a small window; see\cite{BBN} for a similar result. On the other hand the global CKM fits practitioners, see e.g. Refs. 28-30, use  expressions as
\begin{eqnarray}
R_u&=&\left|\frac{U_{ud} U_{ub}^*}{U_{cd} U_{cb}^*}\right|=(1-\frac{\lambda^2}{2})\frac{1}{\lambda}\left|\frac{V_{ub}}{V_{cb}}\right|\\
R_t&=&\left|\frac{U_{td} U_{tb}^*}{U_{cd} U_{cb}^*}\right|=
\frac{1}{\lambda}\left|\frac{V_{td}}{V_{cb}}\right|
\end{eqnarray}
that are not sensitive to such small variations of $a$ and $d$  parameters.
In the second case (\ref{tri3}) both $R$ can take imaginary values and they cannot be ``guessed'' with enough precision for a reasonable fit, and we think that this is the real reason why the second unitarity triangle is not used, although both of them have the same magnitude in $\lambda$\cite{S,F}. However, since  one unitarity triangle provides only two invariant angles,  we need to use  at least two such triangles for the determinations of other two invariant angles for a complete determination of the CKM unitary matrix\cite{AKL}.

\section{Conclusions}

 In the paper we have shown that the unitarity condition method\cite{Di,Di1} and  the unitarity triangles method$^{6-8}$ are both a consequence of the assumed unitarity property for the CKM matrix. In the same time we have reformulated the second method such that the parameters $R_i$ should depend on {\em four} measurable quantities, namely four independent $|U_{ij}|$.

 We have shown that in the real case of approximate experimental data we have to use the gauge invariance property of unitary matrices for imposing all possible constraints implied by the unitarity property. For example the parameterizations
$U_{13}=s_{13} e^{-i\delta}$ and $U_{13}=s_{12}s_{23}e^{-i\delta}-c_{12}c_{23}s_{13}$ have to be treated on the same footing, and have to lead to the same physics.

 We provided  $\chi^2-$tests for both the methods under the form (\ref{chi}) and (\ref{chi-3}), forms that are imposed by the errors that affect all the experimental data, and by the gauge invariance group of unitary matrices.

By Monte Carlo simulations we have shown that the unitarity property requires a very fine tuning between all the CKM matrix elements, and our proposed test  for unitarity implies the fulfillment of two conditions
$\chi^2_1\approx 0$ and $-1 < \cos\delta < 1.$

In our formulation both the methods are exact, they do  not depend on any assumptions concerning the smallness of some parameters and can be used for global fits in both the quark and lepton sectors.

Last but not least we made a clear distinction between the theoretical model given by the matrix $U$, and the experimental data matrix $V$, problem that is  generally overlooked. The unitarity is {\em assumed} and {\em built} in the parametrization of the CKM matrix $U$, but the results of the fits {\em could violate the unitarity} as the Monte Carlo simulations and the numerical examples in section 3 prove; see also\cite{Di}, where the central values published in\cite{HLLL} led  to a few imaginary values for $\cos\delta$. In this sense the most important result for global fits of the CKM matrix entries is the separation criterion between the double stochastic matrices and unistochastic ones, problem that, to my knowledge, was not considered until now in the physical literature, so much the less by the CKM fitting community.

\section*{Acknowledgments}

The author would like to thank C. Alexa and S. Constantinescu for a quick and pedestrian introduction to Monte Carlo simulations. we also thank K. K. Phua for inviting me to submit this contribution to Modern Physics Letters {\bf A}.


\begin{thebibliography}{99}
\bibitem{BBGS}M. Battaglia, A. J. Buras, P. Gambino, and A. Stocchi (Eds.),  Proceedings of the Workshop {\em The CKM Matrix and the Unitarity Triangle},
13-16 February (2002), CERN, Geneva, hep-ph/0304132

\bibitem{AB} H. Abele and D. Mund (Eds.), Proceedings of the Two-Day-Workshop {\em Quark-Mixing, CKM-Unitarity}, September 19-20 (2002), Heidelberg,
hep-ph/0312124

\bibitem{BFKS} P. Ball, J. M. Flynn, P. Kluit, and A. Stocchi (Eds.), Proceedings of the {\em 2nd Workshop on the CKM Unitarity Triangle}, IPPP Durham, April 2003
\bibitem{SD} CKM2005 Workshop, http://ckm2005.ucsd.edu/
\bibitem{Wo} L. Wolfenstein,  Phys.Rev.Lett. {\bf 51} 1945 (1983) 
\bibitem{BLO} A. J. Buras, M. E. Lautenbacher and G. Ostermaier, Phys.Rev {\bf D50} (1994) 3433
\bibitem{CL} M. Ciuchini et al., JHEP {\bf 0701} 013 (2001) 

\bibitem{HLLL} A. H\"ocker, H. Lacker, S. Laplace, and F. R. Le Diberder,  Eur.Phys.J. {\bf C21} 225 (2001) 
\bibitem{Di} P Di\c t\u a, hep-ph/0408013
\bibitem{Ja1} C. Jarlskog, Z. Phys. C {\b9 29}  491 (1985); Phys.Rev.Lett. {\bf 55} 1039 (1985)

\bibitem{Ja2} C. Jarlskog, Phys.Rev. {\bf D 35} 1685 (1987)
\bibitem{BD} J. D. Bjorken and I Dunietz, Phys.Rev. {\bf D 36} 2109 (1987) 
\bibitem{Di1} P. Di\c t\u a, hep-ph/0502125
\bibitem{Ma} L. Maiani, in {\em Proceedings of the International Symposium on Lepton and Photon Interactions at High Energies, Hamburg 1977}, DESY, Hamburg, 1977
 \bibitem{DGW} I. Dunietz, O. W. Greenberg, and D. Wu,  Phys.Rev.Lett. {\bf 55}  2935 (1985)
\bibitem{NP} J. F. Nieves and P. B. Pal,  Phys.Rev. {\bf D 36} 315 (1987)
\bibitem{BB} F. J. Bottela, G. C. Branco, M. Nebot, and M. N. Rebelo, Nucl.Phys. {\bf B 651} 174 (2003)

\bibitem{KM} M. Kobayashi and T. Maskawa, Progr.Theor.Phys. {\bf 49}  652 (1973)
\bibitem{CK} L. L. Chau and W. Y. Keung, Phys.Rev.Lett. {\bf 53} 1802 (1984)
\bibitem{PDG2004} S. Eidelman et al, Phys.Lett. {\bf B 592}  1 (2004)


\bibitem{AKL} R. Aleksan, B. Kayser and D. London,  Phys.Rev.Lett. {\bf 73} 18 (1994) 

\bibitem{MO} A. W. Marshall and I. Olkin, {\em Inequalities: Theory of Majorization and Its Applications}, (Academic Press, New York, 1979), Chapter 2

\bibitem{BL} G. C. Branco and L. Lavoura, Phys.Lett. B {\bf 208} 123 (1988) 
\bibitem{BBN} F. J. Botella {\em et al}, hep-ph/0502133
%\bibitem{PDG2002} K Hagiwara {\em et al.} Phys.Rev. {\bf D 66} (2002) 010001
\bibitem{Je} J. Charles  {\em et al.}, (The CKM Fitter Group), hep-ph/0406184

\bibitem{Bo} M. Bona  {\em et al}, $UT_{fit}$ Collaboration, hep-ph/0501199

http:/www.utfit.org


%\bibitem{HFAG} Heavy Flavour Averaging Group, http://www.stanford.edu/xorg/hfag/triangle/
\bibitem{Br} G. C. Branco and L. Lavoura,Phys.Rev. {\bf D 38} 2295 (1988)
\bibitem{BPS} A. J. Buras, F. Parodi and A. Stocchi, {\bf JHEP 0301} (2003) 029
\bibitem{S} A. Stocchi, hep-ph/0405038
\bibitem{F} R. Fleischer, hep-ph/0405091
\end{thebibliography}
\end{document}